\documentclass[useAMS,usenatbib]{emulateapj}
\shorttitle{The Superorbital Period in LMC X-4}
\shortauthors{Neilsen et al.}

\begin{document}

\title{Spectroscopic Signatures of the Superorbital Period in the Neutron Star Binary LMC X-4}

\author{Joseph Neilsen\altaffilmark{1,2}, Julia C. Lee\altaffilmark{1,2},
  Michael A. Nowak\altaffilmark{3}, Konrad Dennerl\altaffilmark{4},
  Saeqa Dil Vrtilek\altaffilmark{2}} 
\altaffiltext{1}{Astronomy Department, Harvard University, Cambridge,
  MA 02138; jneilsen@cfa.harvard.edu}
\altaffiltext{2}{Harvard-Smithsonian Center for Astrophysics, Cambridge, MA 02138} 
\altaffiltext{3}{MIT Kavli Institute for Astrophysics and Space
  Research, Cambridge, MA 02139}
\altaffiltext{4}{Max-Planck-Institut f\"ur extraterrestrische Physik,
  Giessenbachstra{\ss}e, 85748 Garching, Germany}

\begin{abstract}
We present the first high-resolution X-ray study of emission line
variability with superorbital phase in the neutron star binary LMC
X-4. Our analysis provides new evidence from X-ray
spectroscopy confirming accretion disk precession as the origin of the
superorbital period. The spectra, obtained with the \textit{Chandra}
High-Energy Transmission Grating Spectrometer (HETGS) and the
\textit{XMM-Newton} Reflection Grating Spectrometer (RGS), contain a
number of emission features, including lines from hydrogen-like and
helium-like species of N, O, Ne, and Fe, a narrow O\,{\sc vii} RRC,
and fluorescent emission from cold Fe. We use the narrow RRC and
the He$\alpha$ triplets to constrain the temperature and density of
the (photoionized) gas. By comparing spectra from different
superorbital phases, we attempt to isolate the contributions to line
emission from the accretion disk and the stellar wind. There is also
evidence for highly ionized iron redshifted and blueshifted by $\sim
25,000$ km s$^{-1}$. We argue that this emission originates in the
inner accretion disk, and show that the emission line properties in
LMC X-4 are natural consequences of accretion disk precession. 
\end{abstract}


\section{INTRODUCTION}
The eclipsing binary LMC X-4 consists of a 1.25 M$_{\sun}$
accretion-powered X-ray pulsar orbiting a 14.5 M$_{\sun}$ O8 III star
with a period of 1.4 days (Kelley et al.\ 1983; van der Meer et
al. 2007). The system exhibits a 13.5-s pulsation due to the spin of
the neutron star and a long-period superorbital variation of
approximately 30 d (Lang et al. 1981; Ilovaisky et al. 1984), over
which the X-ray flux may vary by as much as a factor of 60. Similar
behavior has been observed in other X-ray binaries (Her X-1, Giacconi
et al. 1973; SMC X-1, Gruber \& Rothschild 1984).  

The standard model for the superorbital period in LMC X-4 is a precessing
accretion disk that periodically obscures the neutron star \citep{HV89}. For
high-luminosity systems, radiation pressure can generate a stable warp
in the accretion disk, which then precesses due to
radiation torque \citep{P96}. During the low state, an observer views
the edge of the disk and the neutron star is obscured; during the high
state, the disk face is revealed and the neutron star is visible
\citep{P77}. Signatures of this obscuration are well documented in
timing (Her X-1, Petterson et al. 1991; SMC X-1, Hickox et al. 2005;
Neilsen et al. 2004) and broadband spectral (SMC X-1, Vrtilek et
al. 2001; LMC X-4, Woo et al. 1995) studies of pulsar
binaries. Furthermore, high-resolution X-ray studies of the accretion
disk corona in Her X-1 indicate that the disk is in fact observed
edge-on during the low state \citep{JG05}.

There is also substantial evidence to suggest that emission lines in
Her X-1 and LMC X-4 vary with superorbital phase $\Psi$
(Jimenez-Garate et al. 2002, hereafter JG02; Naik \& Paul 2003; Naik
\& Paul 2004). However, the relationship between lines and the
continuum is complex and may require contributions from the accretion
column, the accretion disk, a radiatively-driven stellar wind, or an
X-ray heated wind from the accretion disk atmosphere. The extent to
which these emission lines vary with $\Psi$ depends on the true
physical origin of the superorbital period.

High-resolution spectroscopy is necessary to decompose the spectrum of
LMC X-4 into its physical components. Particularly useful is the
He$\alpha$ emission line triplet (from helium-like ions). This triplet
consists of the closely-spaced resonance line ($r$:
1s$^{2}~^{1}$S$_{0}$ -- 1s2p~$^{1}$P$_{1}$), intercombination lines
($i$: 1s$^{2}~^{1}$S$_{0}$ -- 1s2p~$^{3}$P$_{2,1}$), and the forbidden
line ($f$: 1s$^{2}~^{1}$S$_{0}$ -- 1s2s~$^{3}$S$_{1}$). The ratio
$R=f/i$ can be used to determine the electron density of the emitting
plasma for $10^{8}$ cm$^{-3}<n_{e}<10^{18}$ cm$^{-3},$ and $G=(f+i)/r$
is  sensitive to the electron temperature in the range $10^{6}$ K
$<T_{e}<10^{7}$ K (Gabriel \& Jordan 1969; Porquet \& Dubau
2000). Furthermore, because photo- and collisional excitation populate
atomic levels differently, these line ratios can be used to infer the
dominant excitation mechanism in astrophysical plasmas. 

We have carried out detailed spectral analysis of
the first high-resolution X-ray grating spectra of LMC X-4, obtained with
the \textit{Chandra} HETGS \citep{C05} and the \textit{XMM-Newton}
RGS \citep{dH01}. These observations probe the transition from the low
state to the high state of the superorbital period, revealing the
physical processes at work. We present the first study
of the plasma conditions in LMC X-4 with density and temperature
diagnostics from helium-like species; we exploit the
variability of these lines to constrain the geometry and dynamics of
the line-emitting regions. In \S 2, we describe the observations and
data reduction; in \S 3 we discuss continuum fitting and line
searches. We present and interpret the emission lines from LMC X-4 in
\S 4 and \S 5, and conclude in \S 6. 

\section{OBSERVATIONS AND DATA REDUCTION}

LMC X-4 was observed with the RGS on \textit{XMM-Newton} on 2003
September 9 (10:11:34 UT) for 113.5 ks (approximately one full orbital
period). This observation took place during the high state of the 30 d
superorbital cycle. The transition from low state to high state was
observed with the \textit{Chandra} HETGS on 2007 August 30 (01:52:21
UT), August 31 (10:49:17 UT), and September 1 (16:18:44 UT) for 47.09
ks, 50.96 ks, and 44.06 ks, respectively, for a total exposure time of 
142.11 ks on the source. The observations are reported in Table 1 and
are shown with respect to the superorbital period in Figure 1.

\begin{deluxetable*}{cccccc}
\tabletypesize{\scriptsize}
\tablecaption{Observations of LMC X-4
\label{tbl-1}}
\tablewidth{0pt}
\tablehead{
\colhead{Obs.}  &
\colhead{X-ray}  & 
\colhead{}  &
\colhead{Start}  & 
\colhead{T$_{\rm exp}$}  & 
\colhead{Orbital}  \\
\colhead{ID}  &
\colhead{State}  &
\colhead{Grating}  &
\colhead{Time\tablenotemark{a}}  &
\colhead{(ks)} & 
\colhead{Phase\tablenotemark{b}}}
\startdata
0142800101 & High  & RGS   & 52891.42 & 113.5 & 0.25--1.19  \\
9573       & Transition & HETGS & 54342.08 & 44.1 & 0.28--0.71  \\
9574       & Transition & HETGS & 54343.45 & 51.0 & 0.25--0.62  \\
9571       & Transition & HETGS & 54344.68 & 47.1 & 0.12--0.53
\enddata
\tablenotetext{a}{MJD=JD-2,400,000.5}
\tablenotetext{b}{Computed from the ephemeris of \citet{L00}.}
\end{deluxetable*}

\begin{figure}
\includegraphics[width=3.3 in]{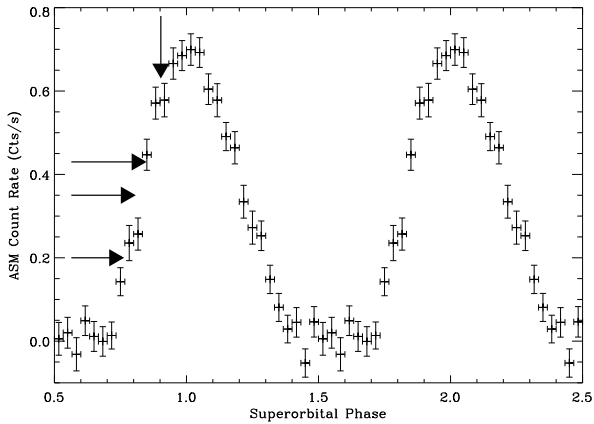}
\caption{RXTE/ASM lightcurve of LMC X-4, folded on the long-term
  period of 30.3 d; two cycles are shown for clarity. $\Psi=1$ is
  defined to be the peak of the high state. The \textit{Chandra} HETGS
  grating observations of the state transition are shown as
  horizontal arrows, and the \textit{XMM-Newton} RGS observation of
  the high state is shown as a vertical arrow.} 
\label{fig1}
\end{figure}

The RGS data were processed with SAS v.\ 7.1.0 and standard data
reduction tools. We obtained spectra from RGS-1 and
RGS-2 with FWHM spectral resolutions of roughly $\Delta\lambda=0.06$
\AA. Bad columns, including CCD chip 7 in RGS-1 and CCD chip 4 in
RGS-2, were flagged and subsequently ignored. We also extracted the
EPIC-PN spectrum in order to constrain the power law index above 3 keV
in the high state. The EPIC-PN spectrum will be analyzed in a future
paper (K. Dennerl, in preparation). For reasons to be discussed in \S
4.1, we also extracted a separate RGS spectrum of LMC X-4 during
eclipse.

The HETGS data were analyzed with the {\sc ciao} analysis suite,
v.\ 3.4.3. After reprocessing and filtering the data, we extracted
Medium Energy Grating (MEG) and High Energy Grating (HEG) spectra,
which have FWHM spectral resolutions of 0.023 \AA~and 0.012 \AA,
respectively. Because the \textit{destreak} tool can cause spectral
artefacts for bright sources, we used the order-sorting routine to
remove the ACIS S-4 streak. To achieve sufficient S/N, we combined all
three HETGS observations. This may involve averaging over real
phenomena, but the third observation is close to an order of magnitude
brighter than the other two combined, so the effect should be small. In
general, we do not account for orbital variability. We performed the
following spectral analysis in ISIS (Houck \& Denicola 2000; Houck
2002). 

\section{CONTINUUM SPECTROSCOPY AND LINE SEARCHES}
Historically, the X-ray continuum in LMC X-4 has been modeled as an
absorbed hard power law plus a soft excess, which is evident because
of the low $N_{H}$ towards the LMC \citep{H04}. The
power law likely consists of Comptonized photons from the neutron
star. Its slope varies between $\Gamma\sim0.5-0.7$; the soft excess has
been described as a combination of thermal Comptonization \citep{LB01},
blackbody radiation \citep{NP02}, and thermal bremsstrahlung
\citep{NP04}. We set the minimum $N_{\rm H}$ at the Galactic value,
$5.78\times10^{20}$ cm$^{-2}$ \citep{DL90} and fit the spectra with
various combinations of these hard and soft components.

\begin{figure}
\includegraphics[angle=270,width=3.3 in]{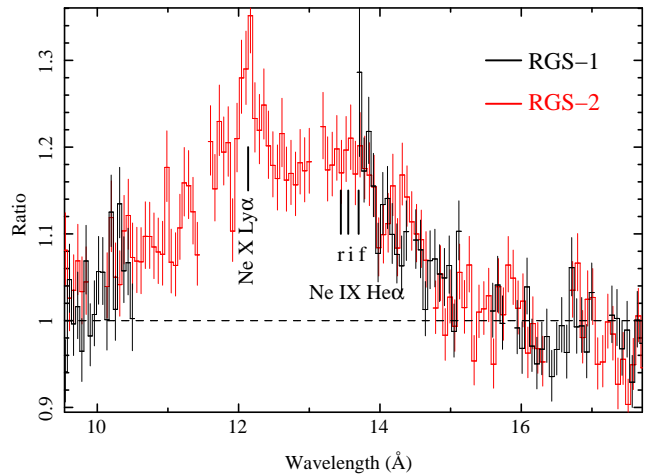}
\caption{Data:model ratio for the RGS spectra near 12 \AA, showing the
  broad bump and the location of emission features from H-like and
  He-like Ne. The bump obscures the Ne\,{\sc ix} lines and prevents a
  completely reliable determination of the Ne\,{\sc x} line flux. In
  our continuum fits, we model it as a broad Gaussian with
  $\Delta\lambda\sim1.3$ \AA~and equivalent width $W_{0}\sim60$ eV.}
\label{fig2}
\end{figure}

However, we are unable to find a completely physical model for the
high-state continuum due to a broad bump near 12 \AA~(see Figure 2;
also see JG02 for a discussion of possible origins for this feature in
Her X-1). This bump is apparent in both RGS detectors. We model it as a
Gaussian with width $\Delta\lambda\sim1.3$ \AA~and 
equivalent width $W_{0}\sim60$ eV. For this reason we will refrain
from drawing any significant physical inferences from our continuum
model. In order to achieve $\chi^{2}_{\nu}\lesssim3.4$ (d.o.f. 565) in
the RGS, it is necessary to include narrow Gaussians to model the
spectral lines. We include ten Gaussian components: six for the
helium-like triplets of nitrogen and oxygen, two for the Ly$\alpha$
lines of nitrogen, oxygen, and two unidentified strong absorption
lines. The bump prevents reliable analysis of the Ne\,{\sc ix} triplet
and the Ne\,{\sc x} Lyman-$\alpha$ line. We let the fluxes of the
Gaussian lines vary when fitting the continuum, then fit all the line
parameters separately. 

Among those tested, the best spectral model for the high state data is
thermal bremsstrahlung with $kT_{\rm bremss}=0.455_{-0.001}^{+0.022}$
keV plus the broad Gaussian bump, the narrow lines, a blackbody with
$kT_{\rm bb}= 43.4_{-0.2}^{+1.7}$ eV and a power law with
$\Gamma=0.813\pm0.007$ (constrained by the EPIC spectrum above 3 keV),
absorbed by the Galactic neutral hydrogen
column ($N_{\rm H}=5.78 \times10^{20}$ cm$^{-2}$). The absorption
lines, which may be interstellar in origin, are found at
23.541$_{-0.030}^{+0.002}$ \AA~and 23.7$_{-1.7}^{+0.4}$ \AA, and have
equivalent widths of $-0.96\pm1.4$ eV and $-0.21_{-0.18}^{+0.19}$ eV,
respectively. This model results in $\chi^{2}_{\nu}=1.55$
(d.o.f. 537), which is not formally acceptable, but there is very
little structure in the residuals and it is not our goal to model
every spectral feature in the RGS.

Because the \textit{Chandra} and \textit{XMM-Newton} observations
probe different X-ray states of LMC X-4, we fit their continuum
parameters separately. Unfortunately, there is insufficient S/N in the
MEG/HEG to constrain the column density independently, so in fitting
the HETGS spectrum we fix $N_{\rm H}$ at the Galactic value (also
measured by the RGS). To ensure that our characterization of the
continuum is accurate given the low S/N, we group the data to a
minimum of 40 counts per bin and include a Gaussian emission line to
account for the strong Fe K$\alpha$ line at 6.4 keV. The best fit for
the HETGS data from 0.5--10 keV is bremsstrahlung with
$kT=0.43\pm0.03$ keV plus a shallow power law with
$\Gamma=0.40\pm0.04.$ The iron line appears at $1.941\pm0.002$ \AA~and
has $W_{0}=130\pm30$ eV; the model gives $\chi^{2}_{\nu}=1.21$ for 886
degrees of freedom. Assuming a distance of 50 kpc to LMC X-4, we
measure unabsorbed continuum luminosities of (0.5--10 keV)
$L_{\rm X}=1.2\times 10^{37}$ ergs s$^{-1}$ in the state transition and
$L_{\rm X}=1.0\times10^{38}$ ergs s$^{-1}$ in the high state.

After finding a satisfactory continuum model for the HETGS, we proceed
to model the spectral lines. We include Gaussian components for the
the triplets from helium-like species of oxygen, neon, and iron as
well as Lyman-$\alpha$ lines from hydrogen-like nitrogen, oxygen,
neon, and iron (these were the strongest lines visible in our
spectra), and find 90\% confidence limits. Because of somewhat low
S/N, we find it necessary to rebin the MEG spectrum to 0.092
\AA~($4\times$ worse than the nominal resolution) for $\lambda>16$
\AA; the rest of the data are binned to the nominal resolution. Our
final fit gives $\chi^{2}_{\nu}=1.08$ for 1758 degrees of freedom. We
show the line parameters in Table 2. The RGS-1 and RGS-2 spectra are
plotted in Figure 3, and the MEG spectrum is plotted in Figure 4. The
RGS eclipse spectrum is plotted in Figure 5.

\begin{figure}
\includegraphics[angle=270,width=3.3 in]{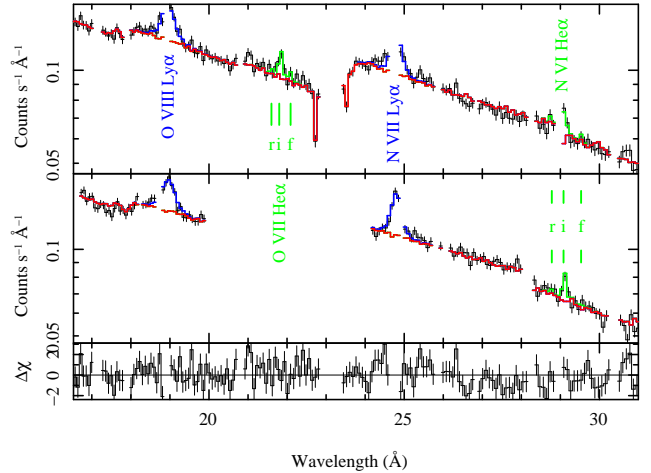}
\caption{RGS high-state spectrum of LMC X-4; RGS-1 is plotted in
  the top panel and RGS-2 is plotted in the center panel. Note the
  strong O\,{\sc viii} and N\,{\sc vii} Ly$\alpha$ lines at 18.97
  \AA~and 24.78 \AA, respectively. Gaps in the spectra correspond to
  ignored columns. The red, green, and blue models correspond
  to the continuum, Region I, and Region II spectra. We show the
  residuals for RGS-1 in the bottom panel.}  
\label{fig3}
\end{figure}

In order to confirm our line IDs, we perform simple checks
with analytic photoionization models from {\sc xstar}, using a power
law with $\Gamma=0.4$ as our ionizing spectrum \citep{K82}. We compare
our spectrum with the {\sc xstar} predictions for other lines and
adjust the model to fit for the appropriate line redshifts and
intensities. In this way we are able to estimate the dynamics and
ionization state of the emitting plasma. In the case of the narrow
iron lines in the HETGS, we find evidence for lines separated from
Fe\,{\sc xxv} by $\sim25,000$ km/s. In order to assess the possibility
that these lines correspond to a physical Doppler shift, we search for
similar Doppler shifts of our other lines as well. This process
results in a total of 21 Gaussian emission lines (stationary or
Doppler-shifted) and two radiative recombination continua (RRC) in the
HETGS.

\begin{deluxetable*}{lccccccc}
\tabletypesize{\scriptsize}
\tablecaption{X-ray Emission Lines in LMC X-4
\label{tbl-2}}
\tablewidth{0pt}
\tablehead{
\colhead{}  & 
\colhead{$\lambda_{0}$}  &
\colhead{$\lambda$}  &
\colhead{$\Delta v_{\rm shift}$}  & 
\colhead{}  & 
\colhead{$W_{0}$}  & 
\colhead{$\sigma_{v}$}  &
\colhead{} \\
\colhead{Line(s)}  &
\colhead{(\AA)}  &
\colhead{(\AA)}  &
\colhead{(km s$^{-1}$)}  &
\colhead{Flux}  &
\colhead{(eV)} & 
\colhead{(km s$^{-1}$)} &
\colhead{Grating}}
\startdata
\multicolumn{8}{c}{Region I: X-ray Irradiated Stellar Wind}\\
Ne\,{\sc ix} He$\alpha$ & 13.447 ($r$) & \nodata & \nodata & $<0.8$ & $1.0_{-0.9}^{+1.3}$ & \nodata & HETG \\
 & 13.552 ($i$) & $13.560_{-0.001}^{+0.004}$ & $190_{-30}^{+100}$ & $1.7_{-0.5}^{+0.6}$ & $5.0_{-1.5}^{+1.8}$ & $<704$ & HETG \\
\vspace{1mm} & 13.698 ($f$) & \nodata & \nodata & $<0.6$ & $<1.7$ & \nodata & HETG \\
O\,{\sc vii} RRC & 16.78 & \nodata & \nodata & $<1.8$ & $1.4_{-1.3}^{+1.7}$ & \nodata & HETG \\
\vspace{1mm}O\,{\sc vii} He$\beta$ & 18.63 & $18.65\pm0.04$ & $280_{-600}^{+580}$ & $3.4_{-1.6}^{+2.1}$ & $4.3_{-2.1}^{+2.6}$ & $700_{-340}^{+550}$ & HETG \\
O\,{\sc vii} He$\alpha$ & 21.602 ($r$) & $21.633_{-0.063}^{+0.007}$ & $440_{-870}^{+100}$ & $6.0_{-2.8}^{+4.1}$ & $5.3_{-2.5}^{+3.6}$ & $<620$ & HETG \\
 & & \nodata & \nodata & $<4.9$ & $<0.4$ & \nodata & RGS \\
 & 21.802 ($i$) & $21.805\pm0.005$ & $40\pm70$ & $13.3_{-4.2}^{+5.5}$ & $ 11.5_{-3.6}^{+4.7}$ & $<400$ & HETG \\
 & & $21.84_{-0.02}^{+0.04}$ & $  470_{-270}^{+500}$ & $12.6_{-3.3}^{+3.4}$ & $0.9_{-0.2}^{+0.3}$ & $<1030$ & RGS \\
 & 22.10 ($f$) & \nodata & \nodata & $<4.0$ & $<3.3$ & \nodata & HETG \\
 & & \nodata & \nodata & $3.7_{- 3.2}^{+ 3.3}$ & $0.3\pm0.2$ & \nodata & RGS \\
N\,{\sc vi} RRC & 22.46 & $22.52\pm0.08$ & $810_{-1040}^{+1080}$ & $<7.6$ & $<6.1$ & \nodata & HETGS \\
N\,{\sc vi} He$\alpha$ & 28.78 ($r$) & \nodata & \nodata & $<6.5$ & $<0.2$ & \nodata & RGS \\
 & 29.08 ($i$) & $29.11\pm0.02$ & $230\pm190$ & $19.4_{-3.7}^{+3.8}$ & $0.6\pm0.1$ & $<370$ & RGS \\\vspace{1mm}
 & 29.53 ($f$) & \nodata & \nodata & $<6.3$ & $<0.2$ & \nodata & RGS \\
\hline
\multicolumn{8}{c}{Region II: Outer Accretion Disk\vspace{1mm}}\\
\vspace{1mm}Fe\,{\sc xxvi} Ly$\alpha$ & 1.781 & $1.785_{-0.012}^{+0.008}$ & $710_{-2100}^{+1280}$ & $<1.4$ & $<33.2$ & \nodata & HETG \\
Fe\,{\sc xxv} He$\alpha$\tablenotemark{a} & 1.850 ($r$) & $1.869_{-0.006}^{+0.009}$ & $3050_{-1060}^{+1510}$ & $3.7_{-1.1}^{+1.2}$ & $86.5_{-25.7}^{+28.6}$ & $3180_{-1120}^{+1560}$ & HETG \\
 & 1.857 ($i$) & \nodata & $1910_{-1050}^{+1500}$ & \nodata & \nodata & \nodata & HETG \\
\vspace{1mm} & 1.868 ($f$) & \nodata & $180_{-1040}^{+1480}$ & \nodata & \nodata & \nodata & HETG \\
\vspace{1mm} Fe K$\alpha$ & 1.938 & $1.941_{-0.002}^{+0.002}$ & $470_{-370}^{+330}$ & $5.8_{- 1.0}^{+1.1}$ & $135.2_{-23.0}^{+25.3}$ & $1200_{-330}^{+430}$ & HETG \\
\vspace{1mm}Ne\,{\sc x} Ly$\alpha$ & 12.135 & $12.158_{-0.003}^{+0.002}$ & $580_{-80}^{+40}$ & $0.9\pm0.3$ & $3.3_{-1.1}^{+1.3}$ & $280_{-200}^{+190}$ & HETG \\
O\,{\sc viii} Ly$\alpha$ & 18.97 & $18.97\pm0.05$ & $-20_{-720}^{+760}$ & $3.3_{-1.8}^{+2.3}$ & $4.1_{-2.2}^{+2.9}$ & $820_{-400}^{+880}$ & HETG \\
\vspace{1mm} & & $18.95\pm0.02$ & $-250\pm280$ & $45.6\pm3.7$ & $4.7\pm0.4$ & $2270_{-290}^{+320}$ & RGS \\
N\,{\sc vii} Ly$\alpha$ & 24.78 & $24.79_{-0.02}^{+0.05}$ & $50_{-190}^{+600}$ & $7.5_{-3.0}^{+4.2}$ & $4.8_{-1.9}^{+2.7}$ & $<330$ & HETG \\
\vspace{1mm} & & $24.79\pm0.02$ & $110\pm180$ & $58.4_{-5.0}^{+5.1}$ & $3.1\pm0.3$ & $1370_{- 180}^{+ 200}$ & RGS \\
\hline
\multicolumn{8}{c}{Region III: Inner Accretion Disk\vspace{1mm}}\\
Fe\,{\sc xxv} He$\alpha$\tablenotemark{a} & 1.850 ($r$) & $1.710_{-0.008}^{+0.006}$ & $-23560_{-1180}^{+ 910}$ & $1.8_{-1.0}^{+1.2}$ & $43.1_{-24.6}^{+29.4}$ & $<4320$ & HETG \\
 & 1.857 ($i$) & \nodata & $-24690_{-1170}^{+900}$ & \nodata & \nodata & \nodata & HETG \\
\vspace{1mm} & 1.868 ($f$) & \nodata & $-26410_{-1160}^{+890}$ & \nodata & \nodata & \nodata & HETG \\
Fe\,{\sc xxv} He$\alpha$\tablenotemark{a} & 1.850 ($r$) & $2.005_{-0.008}^{+0.006}$ & $23950_{-1400}^{+1120}$ & $1.5_{-0.7}^{+0.8}$ & $33.8_{-15.6}^{+17.8}$ & $1910_{-1380}^{+2240}$ & HETG \\
 & 1.857 ($i$) & \nodata & $22820_{-1390}^{+1110}$ & \nodata & \nodata & \nodata & HETG \\ 
\vspace{1mm} & 1.868 ($f$) & \nodata & $21100_{-1380}^{+1100}$ & \nodata & \nodata & \nodata & HETG \\
\vspace{1mm}O\,{\sc vii} He$\beta$ & 18.63 & $20.61_{-0.01}^{+0.07}$ & $30220_{-270}^{+1140}$ & $3.6_{-2.1}^{+3.1}$ & $3.5_{-2.1}^{+3.1}$ & $<930$ & HETG \\
O\,{\sc viii} Ly$\alpha$ & 18.97 & $17.36_{-0.04}^{+0.05}$ & $-26570_{-620}^{+740}$ & $<1.8$ & $<2.8$ & $<5200$ & HETG \\
\vspace{1mm} & & $20.92_{-0.07}^{+0.01}$ & $29190_{-1060}^{+260}$ & $3.1_{-2.2}^{+3.2}$ & $2.9_{-2.1}^{+3.0}$ & $<1100$ & HETG \\
O\,{\sc vii} He$\alpha$ & 21.602 ($r$) & $19.91_{-0.03}^{+0.06}$ & $-24450_{-370}^{+840}$ & $3.0$ & $<3.3$ & \nodata & HETG \\
\vspace{1mm} & 21.802 ($i$) & $20.13_{-0.02}^{+0.09}$ & $-23970_{-110}^{+1250}$ & $2.0_{-1.8}^{+2.6}$ & $2.1_{-1.9}^{+2.8}$ & $<3450$ & HETG \\
N\,{\sc vii} Ly$\alpha$ & 24.78 & $22.91_{-0.06}^{+0.02}$ & $-23470_{-670}^{+130}$ & $8.6_{-5.3}^{+8.7}$ & $6.6_{-4.0}^{+6.7}$ & $<2000$ & HETG 
\enddata
\tablenotetext{a}{Because the Fe\,{\sc xxv} triplet is unresolved, we
  report the Fe\,{\sc xxv} line with three velocities, which
  correspond to the iron velocity assuming the line is dominated by
  the resonance line, the intercombination line, or the forbidden
  line, respectively.} 
\tablecomments{Errors quoted are 90\% confidence ranges for a single
  parameter. $\lambda_{0}:$ rest wavelength; $\lambda:$ measured
  wavelength; $\Delta v_{\rm shift}:$ measured Doppler velocity; Flux:
  measured line flux in units of 10$^{-5}$ photons s$^{-1}$
  cm$^{-2};~W_{0}:$ line equivalent width; $\sigma_{v}:$ measured line
  width. The high state was probed by the RGS, and the transition
  phase was observed with the HETGS (see Figure 1). Notice that the
  lines from Regions I and II are all consistent with a very small
  Doppler velocity (as long as Fe\,{\sc xxv} in the high state is
  dominated by the forbidden line -- see \S 4.3).}
\end{deluxetable*}

\section{EMISSION LINES}
The line emission from LMC X-4 is dominated by hydrogen- and
helium-like ions. In particular, the transition-phase HETGS 
spectrum and the high-state RGS spectrum contain narrow emission lines
from N\,{\sc vi}, N\,{\sc vii}, O\,{\sc vii}, O\,{\sc viii}, 
Ne\,{\sc ix}, and Ne\,{\sc x}. Because of the absence of L-shell iron
emission lines and the presence of RRCs from N\,{\sc vi} and 
O\,{\sc vii} (HETGS only), we conclude that these are recombination
cascade emission lines. As can be seen from Table 2, these lines are
all consistent with little or no Doppler velocity. In the helium-like
triplets, the intercombination ($i$) line dominates. We will discuss
line ratios in \S4.2. In the transition phase we detect a strong
fluorescent Fe K$\alpha$ emission line and recombination lines from
Fe\,{\sc xxv} and Fe\,{\sc xxvi}.   

A comparison of the transition-phase and high-state spectra yields an
interesting discovery: emission lines from helium-like nitrogen,
oxygen, and neon do not evolve in the same way as emission lines from
the corresponding hydrogen-like species. Our analysis leads us to the
conclusion that at least three emission regions are necessary to explain
the line properties in LMC X-4. In the next sections, we will
summarize this argument and discuss each region individually. Because
of chip defects in the RGS (20--24 \AA~and 10.6--13.8 \AA~for RGS-2 and
RGS-1, respectively) and low S/N in the MEG, we will focus on
the O\,{\sc vii} triplet when discussing helium-like species. Note
also that the emission lines are grouped according to region in Table
2. 

\subsection{Line evolution}
The most striking feature of the variable emission lines in LMC X-4 is
that while the emission from N\,{\sc vii} and O\,{\sc viii} is highly
correlated with the continuum flux, the emission from N\,{\sc vi} and
O\,{\sc vii} is not. In particular, while the equivalent widths of the
Lyman-$\alpha$ lines from hydrogen-like N and O appear to be
independent of superorbital phase (at the 90\% confidence level), the
equivalent widths of lines from the corresponding helium-like species
decrease substantially when the continuum flux increases. Interestingly,
the flux in the helium-like lines appears to be constant over
superorbital phase.

\begin{figure}
\includegraphics[angle=270,width=3.3 in]{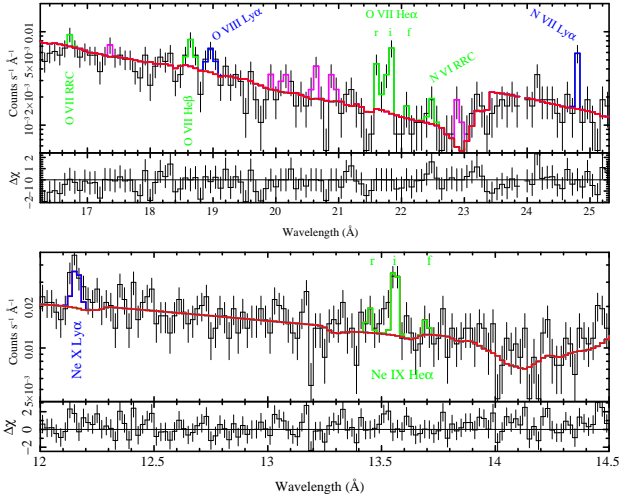}
\caption{MEG transition-phase equivalent of Figure 3 showing data,
  fit, and residuals; again the red, green, blue, and magenta models
  correspond to the continuum, Region I, Region II, and Region III, 
  respectively. We have labeled the strongest lines.}
\label{fig4}
\end{figure}

For example, O\,{\sc viii} Ly$\alpha$ has $W_{0}=4.1_{-2.2}^{+2.9}$ eV
as measured by the HETGS in the transition and $W_{0}=4.7\pm0.4$ eV as
detected by the RGS in the high state. In comparison, the equivalent
width of the O\,{\sc vii} $i$ line decreases by a factor of $\sim13$,
from $W_{0}=11.5_{-3.6}^{+4.7}$ eV in the transition (HETGS) to
$W_{0}=0.9_{-0.2}^{+0.3}$ eV in the high state (RGS). The flux in this
line, however, does not change: we measure (in units of $10^{-5}$
photons s$^{-1}$ cm$^{-2}$) $f=13.3_{-4.2}^{+5.5}$ in the transition
(HETGS) and $f=12.6_{-3.3}^{+3.4}$ in the high state (RGS).

Further comparison of the transition state spectra to the high state
spectra reveals a noticeable increase in the broadening of these
hydrogen-like Ly$\alpha$ lines relative to their helium-like
counterparts: the helium-like triplets have no significant velocity
width in either the transition or the high state. In the high
state, the N\,{\sc vii} and O\,{\sc viii} Lyman-$\alpha$
lines have $\sigma_{v}=1370$ km s$^{-1}$ and  $2270$ km s$^{-1},$
respectively. Because the RGS is able to constrain the width of
narrower lines (N\,{\sc vii} is narrower than 370 km s$^{-1}$ in the
high state), we believe that this broadening is real. If it is thermal
in origin, the implied temperature is $T\sim10^{9}$ K; Doppler
broadening arising from bulk motion is more likely. These values
should be compared to the transition-phase widths, which are $<330$ km
s$^{-1}$ for N\,{\sc vii} and 820 km s$^{-1}$ for O\,{\sc viii}.

\begin{figure}
\includegraphics[angle=270,width=3.3 in]{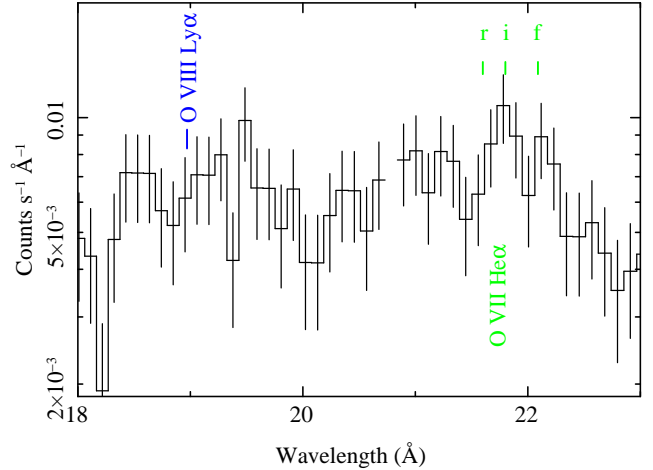}
\caption{RGS-1 eclipse spectrum of LMC X-4. During the eclipse, the
  O\,{\sc viii} Ly$\alpha$ line weakens relative to the O\,{\sc vii}
  triplet, indicating that Region II is not visible during
  eclipse. Given that Region II is highly ionized, exhibits
  somewhat broadened emission lines, and is obscured during eclipse,
  we argue that it is located in the accretion disk.}
\label{fig5}
\end{figure}

These results are difficult to reconcile with a single emission
region with a variable ionization state. The fact that the
hydrogen-like lines brighten by the same factor as the continuum while
the helium-like lines do not change suggests that the hydrogen-like
lines and helium-like lines are produced independently, probably in
two distinct emission regions with different ionization states. The
fact that the O\,{\sc viii} line weakens relative to the O\,{\sc vii}
triplet during eclipse (compare Figures 3 and 5) also supports at
least two emission regions. The presence of high-velocity
($\sim25,000$ km s$^{-1}$) iron during the state transition implies
that a third region may be necessary to explain all the lines in LMC
X-4.

The first region (hereafter Region I), is defined as the source of the
helium-like triplets and weak hydrogen-like lines (note that this
refers only to N, O, and Ne). The luminosity of Region I is not
affected by superorbital variations in the X-ray flux; if it consists
of a photoionized plasma, this might indicate that its luminosity is
tied to the intrinsic luminosity of the neutron star. Region II is the
source of the strong Ly$\alpha$ lines, whose intensity is tied to the
observed X-ray continuum. This region makes little or no contribution
to the helium-like triplets. Region III, for reference, is the origin
of the Doppler-shifted Fe\,{\sc xxv} lines. 

\subsection{Region I: the X-ray irradiated stellar wind?}
As discussed in the preceding section, we define Region I as the
origin of the helium-like emission triplets. These triplets are useful
for plasma diagnostics because their $R$ and $G$ line ratios can
constrain the density, temperature, and dominant ionization process 
in the emitting plasma. Given the striking differences between
line evolution in Regions I and II, these quantities are particularly
important for a complete description of this high-mass binary.

%\subsubsection{$R$ ratio}
In N\,{\sc vi}, O\,{\sc vii}, and Ne\,{\sc ix}, we observe prominent
intercombination lines, weak-to-moderate resonance lines, and weak
forbidden lines; for all these species, $R\sim0$ (with upper limits of
$\sim0.5$ at 90\% confidence). Collisions can produce this line ratio
by exciting the $2~^{3}S_{1}\rightarrow2~ ^{3}P_{1,2}$ transition,
effectively converting forbidden line photons into intercombination
line photons. Thus if collsional ionization dominates, it is possible
to find lower limits on $n$ using the the upper limits on $R$ from the
values in Table 2 and the relationship between $R$ and $n$ determined
by \citet{PD00}. With this method, $R\sim0$ implies a high density
plasma in Region I.

However, photoexcitation by the intense UV field of the secondary can
compete with collisional excitation to produce $R\sim0$
\citep{MS78}. The photoexcitation rate is given by 
\begin{equation}
w_{f\rightarrow i}=\frac{\pi e^{2}}{mc}F_{\nu_{f\rightarrow i}} 
f_{\rm osc},
\end{equation}
where $m$ and $e$ are the mass and charge of the electron, $c$ is the
speed of light, $f_{\rm osc}$ is the oscillator strength, and
$F_{\nu_{f\rightarrow     i}}$ is the flux (in units of photons
s$^{-1}$ cm$^{-2}$ Hz$^{-1}$) at the wavelength of the
$2~^{3}S_{1}\rightarrow2~ ^{3}P_{1,2}$ transition (1637 \AA~for
O\,{\sc vii}). Following JG02, we assume a UV point source (so that
the radiation field and photoexcitation rate $w_{f\rightarrow i}$
decrease with the square of the distance $d$ from the secondary) and
calculate the distance $d_{\rm crit}$ at which the photoexcitation
rate equals the radiative decay rate $w_{f}.$ Inside $d_{\rm crit},$
photoexcitation destroys forbidden-line photons, so the density limits
from $R\sim0$ are not applicable. 

At orbital phase 0.74 (similar to our observations) and superorbital
phase between 0.8 and 1.1 (i.e. in the low-to-high state transition or
in the high state), Preciado, Boroson, \& Vrtilek (2002) measured the
UV flux at 1637 \AA~to be $F_{\lambda_{f\rightarrow i}}
=2.2\pm0.3\times10^{-13}$ ergs s$^{-1}$ cm$^{-2}$ \AA$^{-1}$. Taking
the radiative decay rate and oscillator strength for O\,{\sc vii} from
\citet{D71} and \citet{CT92}, respectively, we calculate $d_{\rm crit}
\sim3\times10^{13}$ cm. This is much larger than the orbital
separation $a=9.3\times10^{11}$ cm \citep{VDM07}, so photoexcitation
cannot be ignored in explaining $R\sim0$. If we neglect collisions
altogether, so that $R\sim0$ is produced by photoexcitation alone, we
find that $R_{\rm I}\lesssim d_{\rm crit}$ (we can ignore the
orbital separation because it is much smaller than 
$d_{\rm crit}$). We conclude that $R$ cannot be used to constrain the
density of Region I. 

%\subsubsection{$G$ ratio}
It should also be possible to constrain the density using the $G$
ratio, because at high densities the resonance line is enhanced by
collisional depopulation of the 1s2s~$^{1}$S$_{0}$ level. This 
effect results in a measureable decrease in $G$ for
$n\gtrsim10^{12}$ cm$^{-3}$ for O\,{\sc vii} \citep{GJ72}. We
attempt to account for collisions by modeling the spectrum with 
{\sc xstar}. We fit the O\,{\sc vii} triplet with a grid of models in
$n$ and $\xi$, covering $1\leq\log\xi\leq2$ and
$\log(n/$cm$^{-3})=\{11,12,13,14,15,16\},$ but find that the
relatively low S/N of the HETGS spectra does not permit a
statistically robust constraint on $n.$

$G$ can also be used to determine the ionization mechanism in Region
I. In the low-density limit, $G=4$ is the cutoff between
photoionized and collisional plasmas; $G>4$ indicates that
photoionization dominates \citep{PD00}. This cutoff is somewhat lower
at high density. Because the $r$ line is a combination of
recombination emission, resonant scattering into the line of sight,
and resonant absorption, $G$ should be interpreted with care
\citep{W03}. This is particularly true during the state transition
(observed with \textit{Chandra}), when the pulsar is at least
partially obscured and resonant scattering into the line of sight is
almost certainly important. In any case, the observed resonance lines
in LMC X-4 are weak, and we are generally only able to place a lower
bound on $G.$ In the high state, we measure $G\gtrsim2.0$ using N and
O; in the transition, we find $G=2.3_{-1.4}^{+4.8}$ and $G>1.5$ for
O\,{\sc vii} and  Ne\,{\sc ix}, respectively. These values allow for
photoionization dominance, but we are are unable to use $G$ alone to
determine the dominant ionization process in LMC X-4. 

However, the comparison of hydrodynamic disk and wind models against
UV spectra has shown that photoionization of the stellar wind can
explain the orbital variability of emission and absorption line
profiles in LMC X-4 (Vrtilek et al. 1997; Boroson et al. 1999; Boroson
et al. 2001). Furthermore, we detect narrow radiative recombination
continua (RRC) from N\,{\sc vi} and O\,{\sc vii}. In the narrow
limit, the width of the RRC is proportional to the electron
temperature. A joint fit to these two features indicates $kT\sim1$
eV, or $T\sim1.2\times10^{4}$ K. The RRC are too faint to provide
confidence limits on $T$ with the {\tt redge} model, but Gaussian fits
(which are acceptable because the features are narrow) imply
$\sigma<4.3$ eV, or $T\lesssim 5\times10^{4}$ K. Because collisional
ionization is negligible at these temperatures, photoionization must
be the dominant process in Region I. For reference, the continuum
emission from Region I falls below the spectral sensitivity of the
HETGS and RGS.

Given these results, we suggest that the X-ray irradiated stellar wind
is a promising candidate for the identity of Region I. We have already
mentioned that this component contributes significantly to the UV
spectrum of LMC X-4, and we have seen that the temperature of Region I
is typical of an O-type star. At this point, we cannot legitimately
distinguish the stellar wind from the face of the secondary. However,
the simplest explanation for the apparent absence of variability on
the superorbital period in Region I is that it occupies a large solid
angle as seen by the neutron star, and the stellar wind meets this
criterion easily. We shall return to this topic in \S 5.

\subsection{Region II: the outer accretion disk/stream?}
Before discussing the Lyman-$\alpha$ lines from N\,{\sc vii}, O\,{\sc
  viii}, and Ne\,{\sc x}, it will prove useful to describe the iron
lines in LMC X-4. During the state transition, we detect a number of
iron emission lines. They are shown in Figure 6, with Gaussian
parameters in Table 2. It is straightforward to identify three of
these lines (at 6.4 keV, 6.7 keV, and 6.9 keV) as the Fe
K$\alpha$ complex, Fe\,{\sc xxv} He$\alpha,$ and Fe\,{\sc xxvi} Ly$\alpha$,
respectively. If we assume that the Fe\,{\sc xxv} emission is
dominated by the forbidden line, then these three iron lines have the
same velocity at 90\% confidence, i.e.\ $v\sim0$ km s$^{-1}$. This is a
reasonable assumption because the critical density for Fe\,{\sc xxv}
is well above $10^{17}$ cm$^{-3},$ so its forbidden line will be
strong for a very wide range of densities \citep{GJ72}. Two 
other lines at 6.1 keV and 7.3 keV constitute Region III and will be
addressed shortly.   

{\sc xstar} photoionization models of this spectral region which
confirm our identification of Fe\,{\sc xxv} He$\alpha$ and 
Fe\,{\sc xxvi} Ly$\alpha$ (see \S 3) indicate $\log\xi\sim3.$ At this 
ionization state, the contribution from helium-like species of N, O,
and Ne is negligible. For the state transition in LMC X-4, {\sc xstar}
predicts Ly$\alpha$ %lines from Ne\,{\sc x}, Si\,{\sc xiv} and
%S\,{\sc xvi} that are roughly consistent with the 1 keV -- 3 keV
%spectrum; 
lines from N and O that are weaker than the Poisson noise in our
spectra. Thus the weak narrow Ly$\alpha$ lines that we detect during
the state transition probably originate in Region I; their strong,
broad counterparts in the high state originate in Region
II. Disentangling the exact contributions to each line from these two
regions is beyond the scope of this paper; full photoionization
modeling will follow in future work.

\begin{figure}
\includegraphics[angle=270,width=3.3 in]{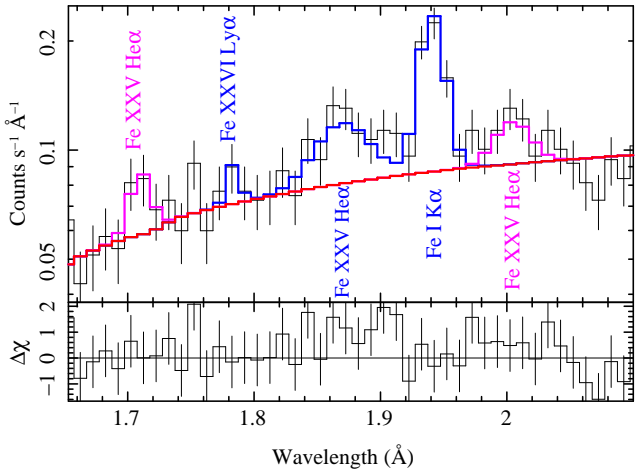}
\caption{HEG transition-phase spectrum of the LMC X-4 iron region. The
  red line is the continuum fit; blue and magenta correspond to
  Regions II and III, respectively.}
\label{fig7}
\end{figure}

At this point, we cannot proceed without associating the
transition-phase iron lines with the high-state Ly$\alpha$ emission
lines from N, O and Ne. This association appears to be confirmed by
the presence of the Fe K$\alpha$ and Fe\,{\sc xxv} lines in the EPIC
high state spectrum (K. Dennerl, in preparation). Furthermore, the
iron line model easily provides the high ionization parameter required
by the absence of helium-like species in Region II. In addition, the
comparable widths of the iron lines and the Ly$\alpha$ lines, along
with their comparable small Doppler velocities, are consistent with a
common origin. Thus we shall treat all these lines as originating in
Region II. Note that the Ly$\alpha$ lines become visible in the high
state because of the correlation between the continuum flux and the
flux from Region II.

Our association of hydrogen-like N, O, and Ne with a hotter region of
the binary also explains the absence of radiative recombination
continua from these species during the high state. At an ionization
parameter $\log\xi\sim3,$ the temperature in Region II is
$T\gtrsim5\times10^{5}$ K ($kT\gtrsim50$ eV). The corresponding RRC
would be very difficult to detect over the bright continuum of the
high state. If the RRC normalizations follow the scaling of
\citet{LP96}, then the absence of a strong O\,{\sc viii} RRC implies
$~T>10$ eV at 90\% confidence. During the state 
transition, there is some evidence for a possible narrow O\,{\sc viii}
RRC with $kT=0.5$ eV ($T\sim6000$ K). Given the above discussion, we
suggest that this feature, if real, actually reflects the contribution
of Region I to the Ly$\alpha$ lines.

Finally, we would like to determine the distance of Region II from the
neutron star. We have an estimate for the ionization parameter in this
region from the iron line model, but no density estimate. However, the
fact that Region II is obscured during eclipse implies a distance from
the neutron star of $R_{\rm II}\lesssim R_{\star}$, where
$R_{\star}=7.8~R_{\sun}$ is the radius of the companion star
\citep{VDM07}. This is a weak constraint, but if we assume that the
line broadening in Region II is due to orbital motion with a Keplerian
velocity profile, then $R_{\rm II}\lesssim1.5\times10^{11}$ cm,
i.e.\ well inside the Roche lobe of the neutron star. We therefore 
conclude that Region II lies in the accretion disk. It should be
pointed out that if the disk is cool and optically thick in the center
but has a hot, optically thin surface layer, then it would be possible
for both Fe\,{\sc xxv} and cold iron (for the Fe K$\alpha$ line) to be
present in Region II. This would also explain their similar velocity 
dispersions and radial velocities but different ionization
states. However, if the width of the Fe K$\alpha$ line is due to a
blend of several charge states, then the cold iron might lie outside
of Region II.

\subsection{Region III: the inner accretion disk?}
In addition to the three iron lines from Region II, we also detected
emission lines at 6.1 keV and 7.3 keV; the {\sc xstar} model for
Region II reveals no obvious candidate IDs at the velocity
of the three strong lines. However, the average of the line energies
coincides with the Fe\,{\sc xxv} He$\alpha$ triplet, so we identify
these lines as symmetrically relativistically Doppler-split 
Fe\,{\sc xxv} emission (with velocities of about 25,000 km
s$^{-1}$). The existence of a number of faint unexplained lines near
other strong emission lines detected at low energy with comparable
velocity shifts (see Table 2) strengthens this interpretation.  

Similar Doppler splitting was observed by EXOSAT during a bright
flare: pulse-phased spectroscopy of the iron line suggested the
existence of two line components whose separation ($\sim 0.2c$) varied
with pulse phase and increased with the luminosity of the source
\citep{D89}. The natural implication of that result is that those
lines originated near the neutron star, possibly in the rapidly
rotating accretion column. This would explain why the Doppler shift
varied with pulse phase, and it is consistent with the visibility of the
neutron star in the high state. In any case, it seems likely that this
motion was dominated by the pulsar's $\sim10^{13}$ G magnetic field
\citep{LB01}.

We consider the possibility that the Doppler splitting in the state
transition is also caused by the pulsar's magnetic field. Because the
iron lines in the transition phase have velocity widths of order 1,000
km s$^{-1}$, we can conclude that they do not originate in the accretion
column (otherwise they would be broadened by $\sim0.2c$ by the spin of the
pulsar). However, material corotating with the magnetic field, e.g. at
the inner edge of the accretion disk, could certainly be responsible
for the observed lines. It is straightforward to calculate that
material corotating with the neutron star at a speed of 25,000 km s$^{-1}$
would lie at $r\gtrsim5\times10^{9}$ cm. This result could place
Region III well inside the accretion disk.

\section{DISCUSSION}
In this paper, we identify a number of recombination emission lines in LMC
X-4; by comparing observations at two epochs, we
determine the evolution of these lines with $\Psi$ and find
evidence for three emitting regions; here we discuss the implications
of these results for the dynamics and geometry of this binary system. 

\subsection{Line variability and emission regions}
Our analysis of the variation of emission lines reveals clear
differences in the properties of helium-like and hydrogen-like N, O,
and Ne in LMC X-4, which we detailed in the previous sections. We
find it useful to suppose that these lines come from two different
physical regions; we will argue that the disk precession model for the
superorbital period, which reproduces the long-term lightcurve,
predicts the existence of these very regions and their
variability. This argument is depicted schematically in Figures 7 and
8, which show the low-state and high-state geometry of LMC X-4 as seen
from the orbital plane.

First, we attempt to place constraints on the physical location of 
Region I. Without a robust density estimate, we cannot use the
ionization parameter to estimate $R_{\rm I}.$ Nevertheless, in \S 4.2
we claimed that we can reasonably identify Region I with the X-ray
irradiated face of the companion star or its substantial wind based on
its variability; we expand on this argument here. In order to conclude
that Region I must occupy a large solid angle as seen by the neutron
star, we must show that shadowing by the precessing disk is unlikely
to produce a constant ionization parameter and line flux over the
superorbital phase.

To do so, we suppose that shadowing is important, i.e.\ that the
illumination of Region I varies like the superorbital lightcurve.
Between the HETGS and RGS observations, the number of ionizing photons
in Region I increases by a factor of $\sim10$, but its ionization parameter
$\xi$ remains constant. The relation $$\xi=\frac{L}{n R^{2}}$$
implies that $nR^{2}$ must increase by a factor of 10 as
well. But given the increase in the number of ionizing photons, Region
I must also be smaller during the high state in order to produce the
same line flux (in the optically thin limit). That is, the neutron
star would have to illuminate a more dense, distant, and smaller
region of the binary during the high state than it does during the
transition.

Clearly shadowing results in a very unusual picture of Region I. The 
accretion geometry is vastly simplified if shadowing is
unimportant and Region I is constantly ionized by the intrinsic
luminosity of the neutron star, i.e.\ $2.3\times10^{38}$ ergs s$^{-1}$
\citep{L00}. In this case, Region I is very likely the X-ray
irradiated wind from the companion. This explanation is
consistent with the low temperature of the emitting plasma as well as
the small line velocities in Region I and the importance of
photoionization in LMC X-4. Furthermore, a relatively steady wind
could provide a solid angle which does not change over time, so that
its ionization parameter and line luminosity in Region I are
independent of $\Psi.$ At present, it is difficult to distinguish
between the surface and the wind of the companion. But the precession
of the disk should cause periodic illumination of the face of the
secondary, so a more complete high-resolution study of line emission
with $\Psi$ might be able to resolve this degeneracy and possibly
constrain the inclination of the disk.

\begin{figure*}
\centerline{\includegraphics[width=6 in]{{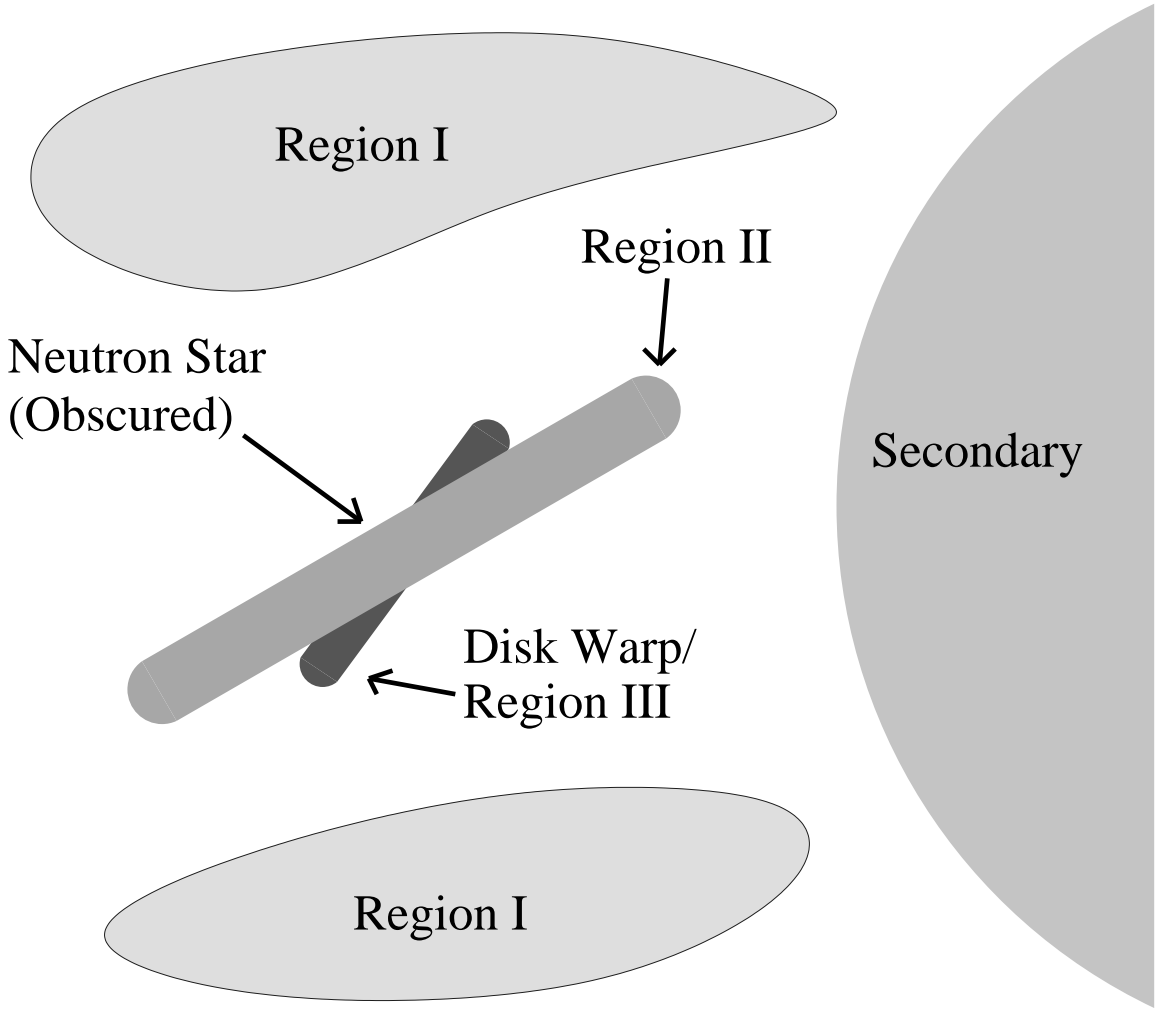}}}
\caption{A cartoon of the X-ray low state of LMC X-4. The disk is
  viewed roughly edge-on, obscuring the neutron star. A warp protrudes
  from the inner disk; as the ionized surface of this warp is dragged
  along by the pulsar's rotating magnetic field, we observe Doppler
  splitting of the disk emission lines. The apparent area of the outer
  disk is small, so lines from Region II are weak. Note that shadowing
  is not particularly important in Region I. This figure should be
  compared to Figure 8, which shows our high-state scenario.}
\label{fig8}
\end{figure*}

\begin{figure*}
\centerline{\includegraphics[width=6 in]{{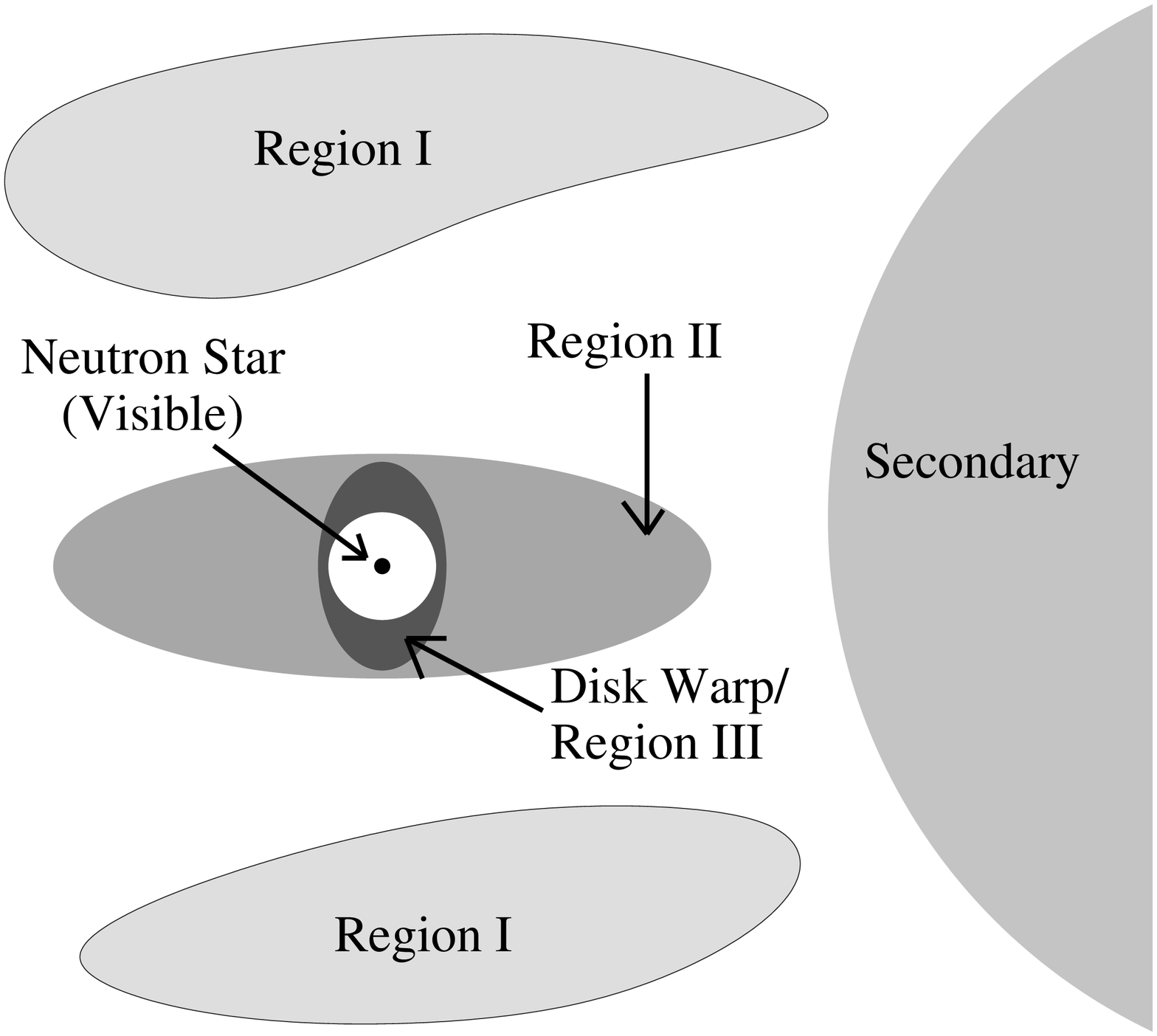}}}
\caption{A cartoon of the X-ray high state of LMC X-4 to be compared
  with Figure 7 (the disk is more face-on now, and the pulsar is
  visible). Because the inclination of the disk has changed, the
  Doppler shifts from the inner disk/Region III are more difficult to
  observe. In contrast, the projected area of the disk
  is much larger, so we observe brighter lines from Region II. The
  solid angle of the wind seen by the neutron star does
  not change, so the line luminosity of Region I is constant. The hole
  in the inner disk represents the magnetosphere, where the magnetic
  field completely dominates the accretion dynamics.}
\label{fig9}
\end{figure*}

While Region I is independent of $\Psi,$ we find that Region II, where
the hot iron lines and the Ly$\alpha$ lines originate, is tied to the
observed X-ray continuum. We argued in \S 4.3 that $R_{\rm II}\lesssim
1.5\times10^{11}$ cm and that Region II probably lies in the outer
accretion disk, which then fills approximately 50\% of the Roche
lobe of the neutron star. Here it is relatively easy to suggest how the observed
equivalent width changes might occur. In the outer disk, the
neutron star is always visible, so the line luminosity is constant in
time (just like in Region I). However, the projected area of the disk
changes periodically, leading to stronger lines when more of the disk
is visible, i.e.\ the high state. It is also possible that the lines
originate in the accretion stream, which is periodically illuminated
by the neutron star so that its line luminosity is tied to the
observed X-ray flux. This effect results in stronger lines during the
high state of the superorbital period. 

JG02 found evidence for a similar increase in emission line
strengths from the low state to the short-on state in Her X-1. They
suggested that the variation in line strengths might be explained by
the enhanced visibility of the disk atmosphere during the
short-on. While it appears that Region II is a common feature of
systems with precessing accretion disks, Region I seems to be absent
in Her X-1. In fact, we can use this point as further evidence that
Region I is the illuminated stellar wind. Her X-1 and LMC X-4 both
have very similar orbital parameters; the main dynamical difference
between these systems is the mass of the companion star
($\sim2.2~M_{\sun}$ in Her X-1 and 14.5 M$_{\sun}$ in LMC X-4). Thus
we expect a stellar wind to contribute much more to the spectrum in
LMC X-4; the absence of a Region I equivalent in Her X-1 is not
surprising if our hypothesis is correct.

It seems that for a system with strong winds, the existence of
two emitting regions is a natural consequence of a tilted,
precessing accretion disk, which illuminates different components of
the binary in different ways. One region presents a roughly constant
solid angle for illumination by the neutron star, so the resulting
emission lines do not vary with $\Psi$. A second component might
precess with the disk or see the neutron star obscured, just as we
do. By assuming the existence of these regions, we can sensibly
decompose our observations into a physical picture of an accreting
neutron star with a precessing accretion disk. 

\subsection{Doppler shifts and the superorbital phase}
We also present evidence for the relativistic Doppler-splitting
of the Fe\,{\sc xxv} He$\alpha$ emission line during the state
transition in LMC X-4, and speculate that the lines originate in the
accretion disk, where highly ionized plasma may be dragged along
with the pulsar's rotating magnetic field at velocities of
$\sim25,000$ km s$^{-1}$. We find evidence of comparably
Doppler-shifted emission lines at low energy which are present during
the state transition but absent in the high state. We consider here
the origin of such Doppler shifts and the implications of the Doppler
evolution for the geometry of the binary.  

The conclusion that the Doppler splitting occurs in the accretion disk
is highly dependent on the velocity field of the disk. For a Keplerian
disk, orbital velocities reach $v\sim25,000$ km s$^{-1}$ at $r_{\rm K}
\lesssim1.9\times10^{7}$ cm. But the disk around a highly
magnetized neutron star is truncated at the radius of the
magnetosphere $R_{\rm M},$ where magnetic pressure balances the ram
pressure of the gas \citep{FKR02}: 
\begin{equation}
R_{\rm M}=1.5\times10^{8}m^{1/7}R_{6}^{-2/7}L_{37}^{-2/7}\mu_{30}^{4/7} 
~{\rm cm}; 
\end{equation} $m$ is the neutron star mass in solar masses, $R_{6}$
its radius in $10^{6}$ cm, $L_{37}$ its luminosity in $10^{37}$
ergs s$^{-1}$, and $\mu_{30}$ its magnetic moment in $10^{30}$ G cm$^{3}.$
With $m=1.25,$ $R_{6}=1$, $L_{37}=23$, and $\mu_{30}=10$ (from
\citealt{LB01}), we find $R_{\rm M}\sim2.3\times10^{8}$ cm. Since the
inner edge of the disk is much larger than $r_{\rm K},$ Keplerian
motion is unlikely to produce the observed Doppler shifts. 

By assuming that Region III is corotating with the pulsar, we
calculate $R_{\rm III}\sim5\times10^{9}$ cm. This is much larger than
$R_{\rm M},$ so the lines cannot originate at the very inner edge of
the disk. However, $R_{\rm III}$ might mark the edge of the inner disk
warp, and a highly ionized surface layer of the disk might continue to
corotate with the magnetic field out to this radius. Thus we conclude
that the Doppler-shifted Region III lies in the inner accretion disk
around LMC X-4.  
 
To clarify this model for visualization purposes, we consider a toy
accretion disk, also shown in Figures 7 and 8. Note that in these
figures, the system is viewed from the orbital plane, not at the
orbital inclination of $\sim68\degr\pm4\degr$ \citep{VDM07}. This disk has a
small but nonzero inclination with respect to the orbital plane, and 
precesses around the orbital angular momentum vector of the
system. The inner disk is warped and slightly tilted with respect to
the disk, so that the edge of the warp protrudes from the disk. The
surface layer of this warp is highly ionized and corotates with the
magnetic field. Furthermore, because only a small part of this warp is
visible the Doppler-shifted lines are narrow, not broad, relative to
the line splitting.

But we have observed LMC X-4 at multiple epochs, and it seems that the
Doppler-shifted lines from Region III may not be visible at all
superorbital phases. In fact, this Doppler evolution is a direct
consequence of the precession of the accretion disk: the changing
orientation of the disk implies a maximum radial velocity in the
accretion disk during the low state, and a minimum in the high
state. Thus Doppler-split lines \textit{should} appear and disappear
at different phases. It seems that we have the first dynamical
confirmation of accretion disk precession as the origin of the
superorbital phenomenon.  

Admittedly, the non-detection of Doppler-shifted lines in the high
state RGS observation does not prove that the disk precessed from
edge-on to face-on. As such, we address the possibilities that the
Doppler splitting is not real (perhaps an effect of low S/N), or that
we observed two distinct phenomena (and not emission lines evolving
with $\Psi$). The Doppler-shifted Fe\,{\sc xxv} lines are both
detected at $>90\%$ confidence, so the conclusion is robust; the
concern over S/N only applies to emission from low-$Z$ species. It is
tempting to suggest that because the fluxes of the Lyman-$\alpha$
lines of hydrogen-like N, O, and Ne are tied to the continuum, the
Doppler-shifted lines should be most detectable in the high state as
well, but there are two reasons why this is not the case. 

First, the projected velocity of the accretion disk is smaller in the
high state; if the disk inclination decreases sufficiently, this effect
could make the Doppler-shifted lines difficult to distinguish from the
stronger $v\sim0$ line. By fitting the RXTE ASM lightcurve and P Cygni
profiles in the UV, \citet{B99} determined a disk tilt of
$31\degr\pm2\degr$ off the orbital plane. Combined with the orbital
inclination of $68\degr$, this result implies
that the apparent inclination of the disk warp varies from $\sim100\degr$
to $\sim35\degr$ over the superorbital phase, so that Doppler velocities
in the high state should be a factor of $\sim2$ smaller than in the
transition. Since our grating spectra can easily resolve such
velocities, the projected velocity of the disk alone cannot explain
the disappearance of the Doppler lines in the high state.

Second, and more importantly, the equivalent widths of the
Doppler-shifted lines should actually be anticorrelated with the
continuum flux. In the high state, only the Fe K$\alpha$
line, the Fe\,{\sc xxv} line, and the Fe\,{\sc xxvi} line are visible
from Region II. As the disk turns its edge to us, the continuum flux
decreases and the Doppler lines appear, boosting the equivalent width
of the iron line region. 

This effect was already observed by \citet{NP03} with the lower spectral
resolution of the RXTE PCA. They found that the equivalent width of
the iron complex was $650_{-36}^{+45}$ eV in the low state, $320\pm30$
eV in the transition, and $280\pm20$ eV in the high state. Our spectra
are consistent with this result: in the transition, we find the
combined equivalent width of the Fe K$\alpha$ line and the
Fe\,{\sc xxv} line to be $220\pm40$ eV; including the Fe\,{\sc xxvi}
line, the equivalent width is $W_{0}\sim 240$ eV. If we include
the Doppler lines, $W_{0}$ increases to $300\pm50$ eV ($\sim 310$
eV when Fe\,{\sc xxvi} is accounted for). Thus the RXTE data are
consistent with our claim that Doppler-shifted lines are undetectable
in the high state, appear in the state transition, and peak during the
low state. High-resolution observations of the low state are necessary
for a definitive test of this interpretation.

It may be that we observed two independent events in the state
transition and the high state, but the observations can be explained
with a single model. The spectroscopic interpretation of the disk
precession model completely explains the variability of the emission
lines in LMC X-4 with superorbital phase; all we require is that the
equivalent widths of the Region II lines be independent of $\Psi$. 

This simple requirement can be met if Region II is in the accretion
disk or otherwise very close to the neutron star. Then the observed
flux in these lines will be strongly correlated with the X-ray
continuum, and their equivalent widths will vary slowly with $\Psi.$
Then as the disk precesses out of the high state, we observe
additional Doppler-shifted emission lines whose velocity and flux
increase into the low state before disappearing again.

\section{CONCLUSION}
Only a handful of XRBs are known to have persistent superorbital
periods, the most famous of which are Her X-1, SS 433, SMC X-1, and
LMC X-4 \citep{W06}. To date, there has been very little systematic
study of the superorbital variation of the strengths or positions of
X-ray emission lines from the precessing accretion disk in any of
these systems (grating studies of SS 433, for example, have focused on
jet knots \citep{L06}). Such studies can be extremely enlightening,
given the power of high-resolution spectroscopy to reveal the plasma
conditions and dynamics in accreting systems.  

We have presented our analysis of time-averaged grating spectra of LMC
X-4 in the high state and state transition of its 30.3 day
superorbital period. In both states, we detected a number of
recombination emission lines; by comparing the intensities and
ionization states of these lines at multiple epochs, we found evidence
for two distinct emission regions, which we identified as the X-ray
photoionized stellar wind (Region I) and the outer accretion disk or
stream (Region II). We argued that the variability of these regions,
which was similar to variability observed in Her X-1 by JG02, was a
natural consequence of accretion disk precession.

We also found evidence for relativistic Doppler splitting of the iron
line in the inner accretion disk, and possible splitting in low-$Z$
recombination lines. We suggested that these lines constitute a third
region, the inner accretion disk warp. Although the Doppler
shifts should be much harder to detect in the high state, the
variation of the equivalent width of the iron line complex with $\Psi$
(as measured by \citet{NP03}) is consistent with our conclusions.

While it is plausible that we simply observed two distinct phenomena
in the different observations, the precession of the accretion disk
provides a simple and coherent explanation for all the phenomena that
we observed. As the disk precesses from edge-on to face-on, the
continuum flux rises; emission lines in the accretion flow brighten
accordingly. At this point, the neutron star is visible and
pulse-phased spectroscopy reveals the rapid motion of the accretion
column, dragged along by the pulsar's magnetic field. As the disk
turns its edge to our line of sight, the neutron star is obscured, but
the radial velocity of the corotating disk is maximized, and we
observe Doppler splitting in the disk emission lines. 

In this work, we have performed a new high-resolution study of X-ray
emission line variability with superorbital phase in LMC
X-4. Accordingly, we are able to report a new spectral 
confirmation of accretion disk precession as the origin of the
superorbital period. The variable Doppler shifts from the inner
accretion disk and the relative evolution of other emission lines in
LMC X-4 are fully consistent with and predicted by the precession of
the accretion disk. Future high spectral resolution observations at
the peak of the high state and during the low state will go a long way
towards a systematic test of this model for accretion dynamics
in X-ray binaries.  

\acknowledgements We gratefully acknowledge funding support
from the \textit{Chandra} Grant GO7-8044X and the Harvard 
University Graduate School of Arts and Sciences. We thank the referee
for useful comments and Jack Steiner for many helpful
discussions. This research has made use of data obtained from the High 
Energy Astrophysics Science Archive Research Center (HEASARC),
provided by NASA's Goddard Space Flight Center.

%{\it Facilities:} \facility{CXO(HETGS)}, \facility{XMM(RGS)}

{}

\label{lastpage}


\begin{thebibliography}{}
%Model of X-ray Photoionization in LMC X-4: Slices of a Stellar Wind
\bibitem[\protect\citeauthoryear{Boroson et al.}{1999}]{B99}
Boroson, B., Kallman, T., McCray, R., Vrtilek, S.~D., \& Raymond,
J.~1999, ApJ, 519, 191

%Testing Hydrodynamic Models of LMC X-4 with Ultraviolet and X-ray
%Spectra 
\bibitem[\protect\citeauthoryear{Boroson et al.}{2001}]{B01}
Boroson, B., Kallman, T., Blondin, J.~M., \& Owen, M.~P.~2001, ApJ, 550, 919 

%The Chandra High-Energy Transmission Grating: Design, Fabrication,
%Ground Calibration, and 5 Years in Flight
\bibitem[\protect\citeauthoryear{Canizares et al.}{2005}]{C05}
Canizares, C.~R., et al.~2005, PASP, 117, 1144

%Oscillator Strengths for S-P and P-D transitions in Heliumlike Ions 
\bibitem[\protect\citeauthoryear{Cann \& Thakkar}{1992}]{CT92}
Cann, N.~M., \& Thakkar, A.~J.~1992, Phys. Rev. A, 46, 5397

%The Reflection Grating Spectrometer on board XMM-Newton
\bibitem[\protect\citeauthoryear{den Herder et al.}{2001}]{dH01}
den Herder, J.~W., et al.~2001, A\&A, 365, 7

%Spectral Variability of LMC X-4
\bibitem[\protect\citeauthoryear{Dennerl}{1989}]{D89}
Dennerl, K.~1989, in Proc. of 23rd ESLAB Symp. on Two Topics in X-Ray
Astronomy, Vol.1: X-ray Binaries, ed. N.E. White, J.J. Hunt, \&
B. Battrick (Paris: ESA), 39 

%H I in the Galaxy
\bibitem[\protect\citeauthoryear{Dickey \& Lockman}{1990}]{DL90}
Dickey, J.~M., \& Lockman, F.~J.~1990, ARA\&A, 28, 215

%Theory of Relativistic Magnetic Dipole Transitions: Lifetime of the
%Metastable 2 3S State of the Heliumlike Ions
\bibitem[\protect\citeauthoryear{Drake}{1971}]{D71}
Drake, G.~W.~F.~1971, Phys. Rev. A, 3, 908

%Accretion Power in Astrophysics
\bibitem[\protect\citeauthoryear{Frank et al.}{2002}]{FKR02}
Frank, J., King, A., \& Raine, D.~2002, Accretion Power in
Astrophysics. (Cambridge: Cambridge Univ. Press)

%Interpretation of Solar Helium-like Ion Line Intensities
\bibitem[\protect\citeauthoryear{Gabriel \& Jordan}{1969}]{GJ69}
Gabriel, A.~H., \& Jordan, C.~1969, MNRAS, 145, 241

%Interpretation of Spectral Intensities from Laboratory and
%Astrophysical Plasmas
\bibitem[\protect\citeauthoryear{Gabriel \& Jordan}{1972}]{GJ72}
Gabriel, A.~H., \& Jordan, C.~1972, in Case Studies in Atomic Physics
II, ed. E.~W. McDaniel, \& M.~R.~C. McDowell (Amsterdam: North-Holland
Publishing Company), 209

%Further X-ray Observations of Hercules X-1 from Uhuru.
\bibitem[\protect\citeauthoryear{Giacconi et al.}{1973}]{G73}
Giacconi, R., Gursky, H., Kellogg, E., Levinson, R., Schreier, E.,
\& Tananbaum, H.~1973, ApJ, 184, 227

%SMC X-1 Variability Observed from HEAO-1
\bibitem[\protect\citeauthoryear{Gruber \& Rothschild}{1984}]{GR84}
Gruber, D.~E., \& Rothschild, R.~E.~1984, ApJ, 283, 546

%Analysis of the optical light curve of the massive X-ray binary LMC X-4
\bibitem[\protect\citeauthoryear{Heemskerk \& van
    Paradijs}{1989}]{HV89}
Heemskerk, M.~H.~M., \& van Paradijs, J.~1989, A\&A, 223, 154

%Origin of the Soft Excess in X-ray Pulsars
\bibitem[\protect\citeauthoryear{Hickox et al.}{2004}]{H04}
Hickox, R.~C., Narayan, R., \& Kallman, T.~R.~2004, ApJ, 614, 881

%Pulse-Phase Spectroscopy of SMC X-1 with Chandra and XMM-Newton:
%Reprocessing by a Precessing Disk?
\bibitem[\protect\citeauthoryear{Hickox et al.}{2005}]{H05}
Hickox, R.~C., \& Vrtilek, S.~D.~2005, ApJ, 633, 1064

%ISIS: An Interactive Spectral Interpretation System for High
%Resolution X-Ray Spectroscopy 
\bibitem[\protect\citeauthoryear{Houck \& Denicola}{2000}]{HD00}
Houck, J.~C., \& Denicola, L.A.~2000, in ASP Conf. Ser. 216, Astronomical
Data Analysis Software and Systems IX, ed. N. Manset, C. Veillet, \&
D. Crabtree (San Francisco: ASP), 591

%ISIS: The Interactive Spectral Interpretation System
\bibitem[\protect\citeauthoryear{Houck}{2002}]{H02}
Houck, J.~C.~2002, in High Resolution X-ray Spectroscopy with
XMM-Newton and Chandra, ed. G. Branduardi-Raymont (Holmbury St. Mary:
Mullard Space Science Laboratory)
http://www.mssl.ucl.ac.uk/\textasciitilde gbr/rgs\_workshop/workshop\_ADS\_ index.html

%LMCX-4 - The optical 30-day cycle and its implications
\bibitem[\protect\citeauthoryear{Ilovaisky et al.}{1984}]{I84}
Ilovaisky, S.~A., Chevalier, C., Motch, C., Pakull, M., van Paradijs,
J., \& Lub, J.~1984, A\&A, 140, 251

%High-Resolution X-ray Spectroscopy of Hercules X-1 with the
%XMM-Newton Reflection Grating Spectrometer: CNO Element Abundance
%Measurements and Density Diagnostics of a Photoionized Plasma
\bibitem[\protect\citeauthoryear{Jimenez-Garate et al.}{2002}]{JG02}
Jimenez-Garate, M.~A., Hailey,C.~J., den Herder, J.~W., Zane, S., \&
Ramsay, G.~2002, ApJ, 578, 391 (JG02)

%Identification of an Extended Accretion Disk Corona in the Hercules
%X-1 Low State: Moderate Optical Depth, Precise Density Determination,
%and Verification of CNO Abundances
\bibitem[\protect\citeauthoryear{Jimenez-Garate et al.}{2005}]{JG05}
Jimenez-Garate, M.~A., Raymond, J.~C., Liedahl, D.~A., \& Hailey,
C.~J.~2005, ApJ, 625, 931

%X-ray Nebular Models
\bibitem[\protect\citeauthoryear{Kallman \& McCray}{1982}]{K82}
Kallman, T.~R., \& McCray, R.~1982, ApJS, 50, 263

%Discovery of 13.5 S X-ray pulsations from LMC X-4 and an orbital
%determination 
\bibitem[\protect\citeauthoryear{Kelley et al.}{1983}]{K83}
Kelley, R.~L., Jernigan, J.~G., Levine, A., Petro, L.~D., \& Rappaport,
S.~1983, ApJ, 264, 568

%The 0.1-100 keV Spectrum of LMC X-4 in the High State: Evidence for a
%High-Energy Cyclotron Absorption Line
\bibitem[\protect\citeauthoryear{La Barbera et al.}{2001}]{LB01}
La Barbera, A., Burderi, L., Di Salvo, T., Iaria, R., \& Robba,
N.~R.~2001, ApJ, 553, 375

%Discovery of a 30.5 Day Periodicity in LMC X-4
\bibitem[\protect\citeauthoryear{Lang et al.}{1981}]{L81}
Lang, F.~L., et al.~1981, ApJ, 246, L21

%Orbital Decay in LMC X-4
\bibitem[\protect\citeauthoryear{Levine et al.}{2000}]{L00}
Levine, A.~M., Rappaport, S.~A., \& Zojcheski, G.~2000, ApJ, 541, 194 

%Photoionization-driven X-ray Line Emission in Cygnus X-3
\bibitem[\protect\citeauthoryear{Liedahl \& Paerels}{1996}]{LP96}
Liedahl, D.~A., \& Paerels, F.~1996, ApJ, 468, L33

%Determining the Nature of the SS 433 Binary from an X-ray Spectrum
%During Eclipse
\bibitem[\protect\citeauthoryear{Lopez et al.}{2006}]{L06}
Lopez, L.~A., Marshall, H.~L., Canizares, C.~R., Schulz, N.~S., \&
Kane, J.~F.~2006, ApJ, 650, 338

%Heliumlike Ion Line Intensities: I. Stationary Plasmas
\bibitem[\protect\citeauthoryear{Mewe \& Schrijver}{1978}]{MS78}
Mewe, R., \& Schrijver, J.~1978, A\&A, 65, 99

%Nature of the Soft Spectral Component in the X-ray Pulsars SMC X-1
%and LMC X-4
\bibitem[\protect\citeauthoryear{Naik \& Paul}{2002}]{NP02}
Naik, S., \& Paul, B.~2002, ApJ, 579, 411 

%Spectral Variations of the X-ray Binary Pulsar LMC X-4 During its
%Long Period Intensity Variation and a Comparison with Her X-1
\bibitem[\protect\citeauthoryear{Naik \& Paul}{2003}]{NP03}
Naik, S., \& Paul, B.~2003, A\&A, 401, 265 

%Timing and Spectral Studies of LMC X-4 in High and Low States with
%BeppoSAX: Detection of Pulsations in the Soft Component
\bibitem[\protect\citeauthoryear{Naik \& Paul}{2004}]{NP04}
Naik, S., \& Paul, B.~2004, ApJ, 600, 351

%Phase Variation in the Pulse Profile of SMC X-1
\bibitem[\protect\citeauthoryear{Neilsen et al.}{2004}]{N04}
Neilsen, J., Hickox, R.~C., \& Vrtilek, S.~D.~2004, ApJ, 616, L135 

%The 35-day Cycle in Hercules X-1
\bibitem[\protect\citeauthoryear{Petterson}{1977}]{P77}
Petterson, J.~A.~1977, ApJ, 218, 783

%A Model for the 35 Day Variations in the Pulse Profile of Hercules
%X-1 
\bibitem[\protect\citeauthoryear{Petterson et al.}{1991}]{P91}
Petterson, J.~A., Rothschild, R.~E., \& Gruber, D.~E.~1991, ApJ, 378,
696 

%X-ray Photoionized Plasma Diagnostics with Helium-like
%Ions. Application to Warm Absorber-Emitter in Active Galactic Nuclei
\bibitem[\protect\citeauthoryear{Porquet \& Dubau}{2000}]{PD00}
Porquet, D., \& Dubau, J.~2000, A\&AS, 143, 495

%Self-Induced Warping of Accretion Disks
\bibitem[\protect\citeauthoryear{Preciado et al.}{2002}]{PBV02}
Preciado, M.~E., Boroson, B., \& Vrtilek, S.~D.~2002, PASP, 114, 340

%Self-Induced Warping of Accretion Disks
\bibitem[\protect\citeauthoryear{Pringle}{1996}]{P96}
Pringle, J.~E.~1996, MNRAS, 281, 357

%Determination of the Mass of the Neutron Star in SMC X-1, LMC X-4,
%and Cen X-3 with VLT/UVES
\bibitem[\protect\citeauthoryear{van der Meer et al.}{2007}]{VDM07}
van der Meer, A., Kaper, L., van Kerkwijk, M.~H., Heemsmerk, M.~H.~M., \&
van den Heuvel, E.~P.~J.~2007, A\&A, 473, 523

%Simultaneous Hubble Space Telescope and ASCA Observations of LMC X-4:
%X-ray Ionization Effects on a Stellar Wind
\bibitem[\protect\citeauthoryear{Vrtilek et al.}{1997}]{V97}
Vrtilek, S.~D., Boroson, B., McCray, R., Nagase, F., \& Cheng, F.~1997,
ApJ, 490, 377

%Simultaneous Chandra and Hubble Space Telescope Observations of SMC
%X-1 
\bibitem[\protect\citeauthoryear{Vrtilek et al.}{2001}]{V01}
Vrtilek, S.~D., Raymond, J.~C., Boroson, B., Kallman, T., Quaintrell,
H., \& McCray, R.~2001, ApJ, 563, L139

%A Systematic Search for Periodicities in RXTE ASM Data
\bibitem[\protect\citeauthoryear{Wen et al.}{2006}]{W06}
Wen, L., Levine, A.~M., Corbet, R.~H.~D., \& Bradt, H.~V.~2006, ApJS,
163, 372

%Resolving the Effects of Resonant X-ray Line Scattering in Centaurus
%X-3 with Chandra
\bibitem[\protect\citeauthoryear{Wojdowski et al.}{2003}]{W03}
Wojdowski, P.~S., Liedahl, D.~A., Sako, M., Kahn, S.~M., \& Paerels,
F.~2003, ApJ, 582, 959

%ROSAT Observations of Scattered X-rays from LMC X-4 in its Low State
\bibitem[\protect\citeauthoryear{Woo et al.}{1995}]{W95}
Woo, J.~W., Clark, G.~W., \& Levine, A.~M.~1995, ApJ, 449, 880

\end{thebibliography}
\end{document}